\documentstyle[prl,twocolumn,aps]{revtex}

\newcommand{\ket}{\rangle}
\newcommand{\bra}{\langle}
\newcommand{\beq}{\begin{equation}}
\newcommand{\eeq}{\end{equation}}
\newcommand{\beqa}{\begin{eqnarray}}
\newcommand{\eeqa}{\end{eqnarray}}
\newcommand{\e}{\mbox{e}}
\newcommand{\w}{\omega}
\newcommand{\wk}{\w_{\mbox{\tiny$_k$}}}
\newcommand{\bk}{{\mbox{\tiny\boldmath$k$}}}
\newcommand{\bkp}{{\mbox{\tiny\boldmath$k'$}}}
\newcommand{\nbk}{{\mbox{\boldmath$k$}}}  
\newcommand{\br}{{\mbox{\small\boldmath$r$} }} 
\newcommand{\nbr}{{\mbox{\boldmath$r$} }}
\newcommand{\brmn}{{\mbox{\small\boldmath$r\mbox{\unboldmath$_{mn}$}$}}} 
\newcommand{\nbrmn}{{\mbox{\boldmath$r\mbox{\unboldmath$_{mn}$}$}}}

\newcommand{\nrmn}{\mbox{\boldmath$r\mbox{\unboldmath$_m-$}r\mbox{\unboldmath$_n$}$}}

\begin{document}
\draft

\title{Decoherence of quantum registers}

\author{John H. Reina$^{1,\ast}$, Luis Quiroga$^{2,\dagger}$ and 
Neil F. Johnson$^{1,\ddagger}$}

\address{$^1$Physics Department, Clarendon Laboratory, Oxford University, 
Oxford, OX1 3PU, UK}

\address{$^2$Departamento de F\'{\i}sica, Universidad de los Andes, A.A. 4976, 
Bogot\'{a}, Colombia}

\twocolumn[\hsize\textwidth\columnwidth\hsize\csname 
@twocolumnfalse\endcsname 
\maketitle 

\begin{abstract}The dynamical evolution of a quantum register of arbitrary length 
coupled to an environment of arbitrary coherence length is predicted within a
relevant model of decoherence. The results are reported for quantum bits (qubits)
coupling individually to different environments (`independent decoherence') and
qubits interacting collectively with the same reservoir (`collective decoherence').
In both cases, explicit decoherence functions are derived for any number of qubits.
The decay of the coherences of the register is shown to
strongly depend on the input states: we show that this sensitivity is a 
characteristic of $both$ 
types of coupling (collective and independent) 
and not only of the
collective coupling, as has been reported previously.  A non-trivial behaviour 
(``recoherence") is found in the decay of the off-diagonal elements of the reduced density matrix
in the specific situation of
independent decoherence. Our results lead to the identification of
decoherence-free states
in the  collective decoherence limit. These states belong to subspaces of the system's
Hilbert space that do not get entangled with the environment, making them ideal
elements for the engineering of ``noiseless'' quantum codes. We also discuss the relations
between decoherence of the quantum register and computational complexity based on the new
dynamical results obtained  for the register density matrix. 
\end{abstract} 



\vskip0.5pc] 
\narrowtext

\section{Introduction}

When any open quantum system, for example a quantum computer, interacts with an 
arbitrary surrounding environment, there are two main effects that
have to be considered when examining the temporal evolution. First, 
there is an expected loss of energy of the initial system due
to its relaxation, which happens at the rate $\tau^{-1}_{rel}$ where $\tau_{rel}$ 
is
the relaxation time-scale of the system.  Second, there 
is a process that spoils the unitarity of the evolution, the
so-called decoherence
\cite{zurek}, where the continuous interaction between the quantum computer and the 
environment leads to unwanted correlations between them in such a way that 
the computer loses its ability to interfere, giving rise to wrong outcomes when
executing conditional quantum dynamics. This phenomenon is characterized by a 
time that defines the loss of unitarity (i.e. the departure of coherence from unity), 
the decoherence time $\tau_{dec}$. 
Usually, the time-scale for this effect to 
take place is much smaller than the one for relaxation, hence 
quantum computers are more sensitive to decoherence processes than to  
relaxation ones. For practical applications in
quantum computing there  are several different systems that might provide a 
long enough $\tau_{rel}$; however, what really matters 
for useful quantum information processing tasks (e.g. quantum algorithms) 
to be performed reliably is the ratio $\tau_{gating}/\tau_{dec}$. 
$\tau_{gating}$, the time required to execute an elementary quantum logic gate,
must be much smaller than $\tau_{dec}$. As a rough estimate, fault-tolerant 
quantum computation has been shown to be possible if the ratio $\tau_{gating}$
to $\tau_{dec}$ of a single qubit is of 
the order of $10^{-4}$ \cite{preskill} (a qubit is a two-state quantum 
system, the basic memory cell of any quantum information processor).

Decoherence and quantum theory are unavoidably connected. Indeed, the ubiquitous 
decoherence phenomenon has been ultimately associated with 
the  ``frontiers" between the
quantum behaviour of microscopic systems and the emergence of the classical 
behaviour observed in macroscopic objects
\cite{zurek}: roughly speaking, the decoherence
time $\tau_{dec}$ determines
the energy and length scales at which quantum behaviour is observed. It 
generally depends non-trivially on several different factors such as 
temperature, dimensionality, quantum vacuum 
fluctuations, disorder, and others whose origin is less well known (hardware 
characteristics). The  time-scale for
decoherence depends on the kind of coupling between the system under  consideration 
and the
environment,  in a range that can go from $pico-$seconds in
excitonic systems \cite{Bonadeo1} up to $minutes$ in nuclear spin systems 
\cite{slichter}.

The discovery of algorithms for which a computer based on the 
principles of 
quantum mechanics \cite{shor} can beat any traditional computer, 
has triggered intense 
research into realistic controllable quantum systems. 
Among the main areas 
involved in this active research field are ion traps \cite{Trapped Ions},
nuclear magnetic 
resonance (NMR) \cite{NMR},  
quantum electrodynamics 
cavities \cite{Cavity QED}, Josephson junctions \cite{Superconductors}, and 
semiconductor quantum dots
\cite{solid}.  The 
main challenge that we face
is to identify a physical system with appropriate internal dynamics and 
corresponding external driving forces which 
enables one to 
selectively manipulate quantum superpositions and entanglements. For this to be done, 
the candidate system should have a sufficiently high level of isolation from the environment:
quantum information processing will be a reality when optimal control  of quantum
coherence in noisy environments can be achieved.  The various 
communities typically rely on different hardware methodologies, and so it is 
important 
to clarify the underlying physics and limits for 
each type of 
physical realization of quantum information processing systems. As mentioned above, 
these 
limits are 
mainly imposed by the 
decoherence time of each particular 
system. However theoretical work has shown that, in addition to 
quantum error correcting codes \cite{shor2,knill}, there are two additional
ways that may potentially lead to decoherence-free or
decoherence-controlled quantum information processing: one of them is the so-called
``error avoiding" approach where, for a given decoherence mechanism
(e.g. collective decoherence discussed below), the existence of ``decoherence-free''
subspaces within the system's Hilbert space  can be exploited in order to obtain a quantum 
dynamics
where the system is effectively decoupled from the environment \cite{dec-free,julia}. This
approach requires the
system-environment coupling to possess certain symmetry properties. 
The other approach is based on ``noise suppression"
schemes, where the effects of unwanted system-bath interactions are dynamically
controlled using a sequence of `tailored external pulses'
\cite{lorenza1,suppression}. These strategies have motivated much 
theoretical and experimental work. In particular, there are some recent
experimental results regarding the engineering of decoherence-free quantum memories
\cite{kielpinski,kwiat}.

We devote this paper to the study of decoherence of an arbitrary quantum register 
(QR) of $L$ qubits. In addition to providing 
a general theoretical framework for studying decoherence, we examine in detail the 
two limits
where  the qubits are assumed  to couple (i) independently 
and (ii) collectively to an external (bosonic) reservoir. The reservoir is modelled
by a continuum of harmonic modes. In Section II we show that the decoherence process 
is very sensitive to the input states of the register and give explicit expressions
for the coherence decay functions. We have found that these functions possess a novel
behaviour which we label ``recoherence'' (or
coherence revivals) in the
specific case of independent decoherence. This behaviour depends strongly on the temperature
and the strength of the qubit-bath coupling.  By contrast, for a certain choice
of the QR input, the calculation of the reduced density matrix leads to the  
identification of decoherence-free quantum states \cite{dec-free,julia} 
when the qubits are coupled ``collectively" to the environment, i.e. when the
distance between qubits is much smaller than the bath coherence length and hence the
environment couples in a permutation-invariant way to all qubits. The  calculations
in this paper were motivated by the results of Ref.
\cite{palma,duan}. The present paper shows that some subtle but important details  of
these earlier results are incomplete. Particularly, the calculation of the $L-$QR density
matrix reported here leads both to new qualitative results, when analyzing the behaviour 
of coherence decay, and new quantitative results, when estimating typical 
decoherence times: these novel results emerge for $L>1$, as discussed 
in Section III. Concluding remarks are 
given in Section IV. We emphasize that the results of this paper are
not restricted
to a particular physical system; they are valid for any choice of the 
qubit system (e.g. photons, 
atoms, nuclei with spin 1/2, etc.)
and any bosonic reservoir.

\section{Independent versus collective decoherence}

Let us consider the general case of a $L-$QR coupled to a quantized environment 
modelled as a
continuum of field modes with corresponding creation (annihilation) bath operator 
$b^\dag$ ($b$). Throughout this work we will analyze 
pure dephasing mechanisms only. We will not consider relaxation mechanisms where the
QR exchanges energy with the environment leading to bit-flip
errors. Hence, the $n^{th}$ qubit operator $J^n\equiv J^n_z$ ($J^n_x=J^n_y=0$)
\cite{nota0}.  As  mentioned earlier, this 
is a reasonable approach since decoherence typically occurs on a much faster time
scale than relaxation. The dynamics of the qubits and the environment can be
described by a simplified version of the widely studied spin-boson  Hamiltonian
\cite{leggett}:

\beq 
H = \sum_{n=1}^L\epsilon_n J_z^n  + \sum_{\bk} \epsilon_{\bk} b_{\bk}^\dag b_{\bk}  
+ \sum_{n,\bk}  J_z^n
(g_{_\bk}^nb_{\bk}^\dag + g_{_\bk}^{n*}b_{\bk} ),
\eeq
where the first two terms describe the
free evolution of the qubits and the environment, and the third term accounts
for the interaction between them. Here $g_{_\bk}^n$ denotes  the  coupling
between the $n^{th}$ qubit and field modes, which in general 
depends on the physical system under consideration. The initial state of the
whole system is assumed to be
$\rho^S(0) = \rho^Q(0) \otimes \rho^{B}(0)$ (the superscripts
stand for system, qubits, and bath), i.e. the QR and the bath are
initially decoupled  \cite{nota1}. We also assume that the environment  is
initially in thermal equilibrium, a condition that can be expressed as
\beq
\rho^{B}(0)=\prod_{\bk} \rho_{_\bk}(T)=\prod_{\bk}\frac{\exp(-\beta\hbar\omega_{_k}b_
{\bk}^\dag
b_{\bk})}{1+\big<N_{\w_{_k}}\big>} \ ,
\eeq
where $\big<N_{\w_{_k}}\big>=\big[\exp(-\beta\hbar\omega_{_k})-1\big]^{-1}$ 
is the Bose-Einstein mean occupation number, $k_B$ 
is the Boltzmann constant and
$\beta\equiv1/k_BT$. For the model of decoherence presented here it is clear 
that we are in a situation where the qubit
operator
$J_z^n$ commutes with the total Hamiltonian $H$ and therefore there is  
no exchange of energy between qubits and environment,  as expected. We 
will not attempt to perform a group-theoretic description for the
study of quantum noise control  \cite{nota2}. Instead, we study the  
specific real-time dynamics of 
the decay in the QR-density matrix coherences
within the model given by the Hamiltonian Eq. (1) (i.e. Abelian noise
processes, in the language of Refs.
\cite{lorenza2,paolo2}).

In the interaction picture, the quantum state of the combined (qubits + bath) 
system at time $t$ is given by
$|\Psi(t) \ket_{_I}=U_{_I}(t)|\Psi(0) \ket$, where $|\Psi(0) \ket$ is the
initial state of the system, $U_{_I}(t)$ is the time evolution operator, 
$U_{_I}(t) = T \exp \big[ -i/\hbar\int_o^t  
H_{_I}(t') dt'\big]$, and $T$ is the time ordering operator. For the Hamiltonian 
(1) we introduce the notation $H=H_0+V$,
where $H_0=\sum_{n=1}^L\epsilon_n J_z^n  + \sum_{\bk} \epsilon_{\bk} b_{\bk}^\dag 
b_{\bk}$ denotes the free evolution term and
$V=\sum_{n,\bk}  J_z^n (g_{_\bk}^nb_{\bk}^\dag + g_{_\bk}^{n*}b_{\bk} )$ is the 
interaction term. Hence, the interaction picture Hamiltonian is given by
$H_{_I}(t)=U_0^\dag(t)VU_0(t)$, with $U_0=\exp(-\frac{i}{\hbar}H_0t)$. A simple 
calculation gives the result
\beq
H_{_I}(t)=\sum_{n,\bk}  J_z^n \Big(g_{_\bk}^n \e^{i \wk t} b_{\bk}^\dag + 
g_{_\bk}^{*n} \e^{-i \wk t} b_{\bk} \Big) \ ,
\eeq
which allows us to calculate the time evolution operator $U_{_I}(t)$. 
The result is (see appendix B.1 for details)
\beq 
U_{_I}(t) = 
\exp\Big[i\Phi_{\w_{_k}}(t)\Big]
\exp \bigg [\sum_{n,\bk}J^n_z 
\left\{ \xi_{\bk}^n(t)b_{\bk}^\dag  - \xi_{\bk}^{\ast n}(t) b_{\bk} \right\}
\bigg]
\eeq
with
\beqa
\nonumber
\Phi_{\w_{_k}}(t) = \sum_{n,\bk}\big|J^n_z g_{_\bk}^n\big|^2 \ 
\frac{\w_{_k}t-\sin({\w_{_k}}t) }{(\hbar{\w_{_k}})^2} \ , \\
\xi_{\bk}^n(t) =
g_{_\bk}^n\varphi_{\w_{_k}}(t)\equiv g_{_\bk}^n \ \frac{1 - \e^{i\wk t}}{\hbar\w_{_k}} \ .
\eeqa
This result differs from the one reported in Ref. \cite{palma} where the 
{\it time ordering
operation for 
$U_{_I}(t)$ was not performed}. As will become clear later, this correction alters 
the resulting calculation of Ref. \cite{palma}, and hence changes the results
for the reduced density matrix of an arbitrary $L-$QR. Based on the time evolution
operator of Eq. (4), we report here
a different result for this density matrix and discuss its implications 
with respect to those of Ref.
\cite{palma}. 

Unless we specify the contrary, we will assume that the coupling coefficients
$g_{_\bk}^n$ ($n=1,...,L$) are {\it position-dependent}, i.e. that each qubit couples 
individually to a different heat bath, hence the
QR decoheres `{\it independently}'.  Implications for the `collective' decoherence 
case will be discussed later. Let us assume that the
$n^{th}$ qubit experiences a coupling to the reservoir characterized by
\beq
g_{_\bk}^n=g_{_\bk} \exp(-i \mbox{\boldmath$k\mbox{\unboldmath$\cdot$} 
r\mbox{\unboldmath$_n$}$}) \ ,
\eeq
where $\nbr_n$ denotes the position of the $n^{th}$ qubit. It is easy to see that the
unitary evolution operator given by Eq. (4) produces entanglement between register 
states and environment states
\cite{palma}. The degree of the entanglement  produced will strongly depend on the 
qubit input states  and also on the separation $||\nrmn||$  between qubits
because of the position dependent coupling. As we will 
show later, for some kind of input states no decoherence occurs at
all despite the fact that all of the qubits are interacting with the environment. 
We will also identify states with 
dynamics decoupled from thermal fluctuations; this fact may be relevant 
when designing experiments where the involved
quantum states have dephasing times (mainly due to temperature dependent effects) 
on a very short time scale, as in the
solid state for example. We will also see that the above features are key issues 
when proposing schemes for maintaining
coherence in quantum computers \cite{dec-free}.  

Due to the pure dephasing (i.e. Abelian) characteristic of the noise model considered
here, we can calculate analytically the functional dependence of the decay  of the
coherences of the QR by taking into account all the field modes of the quantized
environment.  We shall  follow the notation of Ref. \cite{palma}.
Let us consider the reduced density matrix of the $L-$QR: the matrix 
elements of this reduced operator can be
expressed as
\beq
\rho_{_{\{i_n,j_n\}}}^Q(t)  = 
\bra i_{_L},i_{_{L-1}},... , i_{_1} | Tr_{_B}\{\rho^S (t)\}| j_{_L},j_{_{L-1}},...,
j_{_1}\ket \ , 
\eeq
where $\{i_n,j_n\}\equiv\{i_1,j_1;i_2,j_2;...;i_{_L},j_{_L}\}$ refers to the 
qubit states of the QR and 
\beq
\rho^S (t)=
U_{_I}(t)\rho^Q(0)
\otimes \rho^{B}(0)U_{_I}^\dag(t) \ . 
\eeq
From this expression it becomes clear that the dynamics of the register is
completely determined by the evolution operator $U_{_I}(t)$. In Eq. (8), the 
initial density matrix of the register can be
expressed as $\rho^Q(0) = \rho^Q_{i_{_1},j_{_1}}(0)\otimes\rho^Q_{i_{_2},j_{_2}}(0)
\otimes...\otimes\rho^Q_{i_{_L},j_{_L}}(0)$,
where $\rho^Q_{i_{_n},j_{_n}}= | i_{_n}\ket \bra j_{_n}|$. In this expansion, 
$| i_{_n}\ket=|\pm \frac{1}{2}\ket$ are the
possible  eigenstates of
$J^n_z$ and will be associated with the two level qubit states $|1\ket$ and $|0\ket$, 
respectively. The eigenvalues 
$i_{_n}=\pm \frac{1}{2}$ and $j_{_n}=\pm \frac{1}{2}$. In what follows, the subscripts 
of Eq. (7) will be renamed with the
values 1 and 0 to indicate the actual values $\frac{1}{2}$ and $-\frac{1}{2}$, 
respectively. Hence we can rewrite Eq. (7) as
\beq
\rho_{_{\{i_n,j_n\}}}^Q(t)  = Tr_{_B}\Big[\rho^B (0)U^{\dag \{j_{_n}\}}_{_I}(t)
U^{\{i_{_n}\}}_{_I}(t)\Big]\rho^Q_{_{\{i_n,j_n\}}}(0) \ ,
\eeq
\beqa 
\nonumber
\hspace{-0.6cm}
U^{\{i_{_n}\}}_{_I}(t)  \equiv && 
\exp\Big[i \sum_{\bk}\big|g_{_\bk}\big|^2 s(\w_{_k},t)
\sum_{m,n}\e^{i{\mbox{\small\boldmath$k$}}\cdot
\brmn}i_mi_n\Big]\times 
\eeqa
\vspace{-0.6cm}
\beq
\exp \Big[\sum_{n,\bk}\left\{g_{_\bk}\varphi_{\w_{_k}}(t) 
i_ne^{-i{\mbox{\small\boldmath$k$}} \cdot \br_n} b_{\bk}^\dag  - g^\ast_{_\bk}
\varphi^\ast_{\w_{_k}}(t) 
i_n \e^{i{\mbox{\small\boldmath$k$}} \cdot \br_n} b_{\bk} \right\}\Big],
\eeq
and calculate explicitly the decay of the coherences for the $L-$QR. The result 
is (see appendix B.2)
\beqa 
\hspace{-1.45cm}
\rho_{_{\{i_n,j_n\}}}^Q(t)  & = & 
\exp\Big[-\sum_{\bk;m,n}\big|g_{_\bk}\big|^2 
c(\w_{_k},t) \ \times
\eeqa
\vspace{-0.65cm}
\beqa
\hspace{-1.3cm}
\coth\Big(\frac{\hbar\w_{_k}}{2k_{_B}T}
\Big)(i_m-j_m)(i_n-j_n)\cos\nbk\cdot\nbrmn\Big]\times \nonumber 
\eeqa
\vspace{-0.6cm}
\beqa
\hspace{-0.5cm}
\nonumber
\exp \Big[i \sum_{\bk;m,n}\big|g_{_\bk}\big|^2 s(\w_{_k},t)(i_mi_n-j_mj_n)
\cos\nbk\cdot\nbrmn\Big]\times 
\eeqa
\vspace{-0.65cm}
\beqa
\hspace{-0.1cm}
\nonumber
\exp \Big [-2i\sum_{\bk;m,n}\big|g_{_\bk}\big|^2 c(\w_{_k},t)i_mj_n\sin\nbk
\cdot\nbrmn\Big]\rho_{_{\{i_n,j_n\}}}^Q(0), 
\eeqa
where $\nbrmn\equiv\nrmn$ is the relative distance between the $m^{th}$ 
and $n^{th}$ qubits,
$s(\w_{_k},t)=[\w_{_k}t-\sin(\w_{_k}t)]/(\hbar\omega_{_k})^2$, 
and $c(\w_{_k},t)=[1-\cos(\w_{_k}t)]/(\hbar\w_{_k})^2$.  In the continuum
limit, Eq. (11) reads
\beqa
\nonumber
\hspace{-0.85cm}
\rho_{_{\{i_n,j_n\}}}^Q(t) & = &  
\exp\Big[i\big\{\Theta_{d}(t,t_s)-\Lambda_d(t,t_s)\big\}\Big]\times \\ &&
\exp\Big [-\Gamma_d(t,t_s;T)\Big] \ 
\rho_{_{\{i_n,j_n\}}}^Q(0) \ ,
\eeqa
where
\vspace{-0.2cm}
\beqa
\nonumber 
\hspace{-0.95cm}
\Theta_d(t,t_s) & = & \int d\w I_d(\w) s(\w,t)\times
\\ &&
2\sum_{m=1,n=2 \atop m\neq
n}^L(i_mi_n-j_mj_n)\cos(\w t_s) \ ,
\eeqa
\vspace{-0.5cm}
\beq
\hspace{-0.5cm}
\Lambda_d(t,t_s) = 2\int d\w I_d(\w)
c(\w,t)\sum_{m=1 \atop n=2}^Li_mj_n\sin(\w t_s) \ ,
\eeq
and
\vspace{-0.2cm}
\beq
\hspace{-1.1cm}
\Gamma_d(t,t_s;T) = \int d\w I_d(\w) c(\w,t)\coth\Big(\frac{\w}{2\w_{_T}}
\Big)\times
\eeq
\vspace{-0.8cm}
\beqa
\nonumber
\Big\{\sum_{m=1}^L(i_m- 
j_m)^2 
+2\sum_{m=1,n=2 \atop m\neq
n}^L(i_m-j_m)(i_n-j_n)\cos(\w t_s)\Big\}.
\eeqa
Here $\w_{_T}\equiv k_{_B}T/\hbar$ is the thermal frequency 
(see discussion below), 
$\w t_s\equiv \nbk\cdot\nbrmn$ sets the {\it ``transit time"} $t_s$,
and
$I_d(\w)\equiv\sum_\bk\delta(\w-\w_{_\bk})|g_{_\bk}|^2\equiv \frac{dk}{d\w}G(\w)|g(\w)|^2$ 
is the  spectral density of the bath. This function is characterized by a cut-off 
frequency  $\w_c$ that depends on the particular physical 
system under consideration and sets $I_d(\w)\mapsto0$ for $\w>>\w_c$
\cite{leggett}.
We see that an explicit calculation for the decay of the coherences requires 
the knowledge of the spectral density
$I_d(\w)$.   Here we assume that $I_d(\w)=\alpha_d\w^d\e^{-\w/\w_c}$ \cite{leggett}, where 
$d$ is the dimensionality of the field and $\alpha_d>0$ 
is a  proportionality constant that characterizes the strength of
the system-bath coupling. Hence, 
the functional dependence of the
spectral density relies on the dimensionality of the frequency dependence of 
the density of states $G(\w)$ and of the coupling
$g(\w)$.  In Eq. (12) 
the ``transit time" $t_s$ has been introduced in order to express the QR-bath coupling in
the  frequency domain. This transit time is of particular importance when describing the
specific ``independent" decoherence mechanism, because of the position-dependent coupling
between qubits and bath modes. However, in a scenario where the
qubits are coupled ``collectivelly" to the same  environment, $t_s$ does not play any role
(see below).

The result of Eq. (12) differs in several respects from the one 
reported in 
\cite{palma}: the decoherence function $\Gamma_d(t,t_s;T)$ 
contains additional information about the characteristics of the 
independent 
decoherence and the way  in which the individual qubits couple to 
the environment through the 
position-dependent terms which are proportional to $\cos(\nbk\cdot\nbrmn)$. 
In essence this means that 
the entanglement of the register with the noise field depends on the 
qubit separation. 
Also the expression for
$\rho_{_{\{i_n,j_n\}}}^Q(t)$  reveals the new dynamical factor 
$\aleph_{d}(t,t_s)\equiv 
\Theta_d(t,t_s)-\Lambda_d(t,t_s)$ 
which must be taken into account when determining typical
decoherence times for the $L-$QR.  

It is interesting that the
decoherence effects arising from thermal noise can be separated from the 
ones due purely to vacuum fluctuations.
This is simply because the average
number of field
excitations at
temperature $T$ can be
rewritten as $\big<N_\w\big>_T=\frac{1}{2}\exp(-\hbar\w/2k_{_B}T)
$cosech$(\hbar\w/2k_{_B}T)$, and hence 
$\coth
(\hbar\w/2k_{_B}T)=1+2\big<N_{\w}\big>_T$ in Eq. (15).
The other term contributing 
to the decay of the coherences in Eq. (12), $\aleph_{d}(t,t_s)$, is due purely to 
quantum vacuum
fluctuations. The separation made above allows us to examine the different time-scales 
present in the (QR + bath) system's dynamics. The fastest time scale of the
environment is determined by the cut-off:
$\tau_c 
\sim \w_c^{-1}$, i.e. $\tau_c$ 
sets the ``memory" time of the environment.  Hence, the vacuum fluctuations will 
contribute to the dephasing process only for times $t>\tau_c$.
Also note
that the characteristic thermal frequency $\w_{_T}\equiv k_BT/\hbar$ sets another 
fundamental time
scale
$\tau_{_T}
\sim \w_{_T}^{-1}
$: thermal effects will affect the qubit dynamics only for $t >  \tau_{_T}$. Hence 
we see that quantum vacuum
fluctuations contribute to the dephasing process only in the regime $\tau_c <t <\tau_{_T}$. 
From this identification it becomes clear that the qubit dynamics and
hence the decoherence process of our open quantum system will depend on 
the ratio of the temperature-to-cut-off parameter 
$\w_{_T}/\w_c$ and the spectral function $I_d(\w)$. Later we will 
analyze how different
qualitative behaviours are obtained for the decoherence depending on the 
relationship between the cut-off 
and the thermal frequency. It is worth noticing that the analytical separation between
the thermal and vacuum contributions  to the overall decoherence process is a
convenient property of the pure dephasing (Abelian) model considered here. Further
generalizations of this model, e.g. by including relaxation mechanisms, should make this
separation non-trivial because the problem becomes no longer exactly solvable.

Next, we analyze the case of ``collective decoherence".
This situation can be thought of as a bath of ``long" coherence length
(mean effective wave length $\lambda$) if compared with the separation 
$r_{mn}$ between 
qubits, in such a way that  $\lambda>>r_{mn}$ and hence the product of
Eq. (6) has
$\exp(-i\nbk\cdot\nbr_{n})\approx 1$. Roughly speaking, we are in a situation where
all the qubits ``feel''  the same environment, i.e.  $g_{_\bk}^n\equiv g_{_\bk}$.
A similar  calculation to the one
followed in Appendix B gives the following result for the decay of the 
coherences for the case of collective coupling to the reservoir:
\beqa
\nonumber
\rho_{_{i_n,j_n}}^Q(t) & = &
\exp\bigg\{i \Theta_{d}(t)\bigg[\bigg(\sum_{m=1}^Li_m\bigg)^2-\bigg
(\sum_{m=1}^Lj_m\bigg)^2\bigg]
\bigg\}\times \\ &&
\exp\bigg\{-\Gamma_d(t;T)\bigg[\sum_{m=1}^L\big(i_m-j_m\big)\bigg]^2
\bigg\} \ 
\rho_{_{i_n,j_n}}^Q(0) \nonumber \\
\eeqa
where
\vspace{-0.2cm}
\beq 
\hspace{-1.5cm}
\Theta_d(t) = \int d\w I_d(\w) 
s(\w,t) \ ,
\eeq
and
\vspace{-0.2cm}
\beq
\nonumber
\Gamma_d(t;T) = \int d\w I_d(\w)c(\w,t)\coth\Big(\frac{\w}{2\w_{_T}}
\Big) \ .
\eeq
Expressions (12) and (16) are to be compared with those reported in Refs. \cite{palma} and 
\cite{duan}. Clearly, the new result for the evolution operator
$U_I(t)$ of Eq. (4) induces a QR-environment dynamics different to the one reported in 
\cite{palma}. This fact has been pointed out in Ref. \cite{duan} for the
situation of collective decoherence; in this particular case, our results coincide with
theirs. However,  we obtain different results when considering the situation of
independent decoherence: the new dynamical factor
$\aleph_{d}(t,t_s)$ includes additional information about individual qubit dynamics that
cannot  be neglected. Indeed, 
from the results derived in Ref. \cite{duan}, the authors argue that the
damping of the independent decoherence is insensitive to the type of initial states and 
hence 
the sensitivity to the input states is only a property of the
collective decoherence. As can be deduced from Eqs. (12) and (16), 
we find that this statement is not generally true and that the sensitivity to the
initial states is a property of both collective $and$ independent
decoherence. This result is  particularly illustrated for the case of a $2-$QR in the next
section. From the expressions (17) and (18) we see that for $\hbar\omega<<k_BT$, 
the high-temperature environment (high-TE), the phase damping factor
$\Gamma_d(t;T)$ is the main agent responsible for the qubits' decoherence while the other 
dynamical damping factor $\Theta_d(t)$ plays a minor role. In this case $\w_c$ is actually 
the only characteristic frequency accessible to the system ($\w_c<<\w_{_T}$) and thermal 
fluctuations always dominate
over the vacuum ones.
However, when we consider the situation $\w_c>>\w_{_T}$, the low-temperature environment (low-TE), 
these damping factors compete with each other over the
same time scale, and $\Theta_d(t)$ now plays a major role in eroding the qubits' coherence. 
Here there is a much more interesting dynamics between thermal and vacuum contributions 
(see next section). This
shows the difference and the importance of the additional terms of the reduced density matrix 
reported here when compared with those of Refs. \cite{palma,duan}.
The above statements will be illustrated in the
next section. 

So far, the dynamics of the qubits and their coupling to the environment has been
discussed  in the interaction representation. To go back to the
Schr\"odinger representation, recall 
$\rho_{_{Sch}}(t) =  U_{_0}(t)\rho_{_{I}}(t)U^{\dag}_{_0}(t)$, with $\rho_{_{I}}(t)$ as calculated before for the 
qubits decoherence  (Eqs. (12) and (16)). Also note that in the
Schr\"odinger representation (here denoted with subindex $Sch$) $U_{_0}(t)$ introduces mixing 
but not decoherence.
Next, we consider some particular cases which allow us to evaluate the expressions (12) 
and (16) and hence give a qualitative
picture of the respective decoherence rates for both collective and independent decoherence situations.

\section{Dimensionality of the field and decoherence rates for few qubit systems}

In this section we analyze the qualitative behaviour of 
the decay of the coherences given by Eqs. (12) and (16) for single 
and two qubit systems.

\subsection{Single qubit case}

Here we consider the case of only one qubit in the presence of a thermal reservoir, 
as defined in Eq. (2). Hence, for both types of coupling (Eqs. (12) and (16)) we get
\beqa
\rho_{_{i_n,j_n}}^Q(t) & = &  \exp[-\Gamma_d(t,T)]\rho_{_{i_n,j_n}}^Q(0)\ .
\eeqa
By using Eq. (4), with $d=1$, it is easy to show that the populations remain unaffected: 
$\rho_{_{ii}}^Q(t)=\rho_{_{ii}}^Q(0)$, $i=0,1$.
In general ($i\neq j$), the decay of coherences is determined by Eq. (19). Here we can
identify the  main time regimes of decoherence for different dimensionalities of
the field. In what follows, we consider the case of reservoirs with both  
one-dimensional density
of states (``Ohmic") 
and  
three-dimensional density of states (``super-Ohmic"), i.e. 
$G(\w)=constant$ and
$G(\w)=\w^2$, respectively, where the frequency dependent coupling
$g(\w)\propto\sqrt\omega$, as considered in \cite{palma}. From Eq. (19) we see that
a general solution for the case $d=1$
requires numerical integration (see Appendix A). However, for the case where the interplay 
between thermal and vacuum effects is more complex, i.e. the low-TE
($\w_{_T}<<\w_c$), we can solve it  
analytically. Here we get
\beqa
\nonumber
\Gamma_1(t,T) &  \approx & c_1 \bigg[ 2\omega_{_T}t\arctan (2\omega_{_T}t)  + 
      \frac{1}{2}\ln  \bigg( \frac{1 + \w_c^2t^2}{1 + 4\omega_{_T}^2t^2} 
   \bigg) \bigg] \ , \\ &&
\eeqa
where $c_1\equiv\alpha_1/\hbar^2$. On the other hand, an exact solution 
for the super-Ohmic case 
$d=3$ (Eq. (19))
 can be found for any temperature $T$. The result is: 
\beqa 
\nonumber
\Gamma_3(t,T) & = & c_3\bigg\{ \theta^2\Big[
\zeta(2,\theta) + 
    \zeta(2,1+\theta)-
    \zeta(2,\theta+i\omega_{_T} t)-  \\ &&
    \zeta(2,\theta-i\omega_{_T} t)\Big] +
    \frac{1-\w_c^2t^2}
      {[1+\w_c^2t^2]^2}\bigg\} \ ,
\eeqa
where $\zeta(z,q)=[1/\Gamma(z)]\int _{o}^{\infty} dt \ 
[t^{z-1}\e^{-qt}/(1-\e^{-t})]$ is the generalized Riemann zeta function,
$\Gamma(z)=\int _{o}^{\infty} dt \ t^{z-1}\e^{-t}
$ is the Gamma function, and  $c_3=\alpha_3\w_c^2/\hbar^2$. 

For the purpose of this paper, we concentrate 
on the case $L>1$ for which we have several novel results. 
We leave the analysis of the
$L=1$ case to Appendix A, where we discuss the process of identifying 
typical decoherence times for a single qubit, and the interplay between 
the different decoherence 
regimes as a function of the temperature.

\subsection{$L=2$ qubit register}


Let us analyze the case of two qubits in the presence of the bosonic 
reservoir discussed in the present paper. Using the same expressions 
for the density
of states $G(\w)$ and for the frequency dependent coupling
$g(\w)$ as above,  we will analyze the coherence decay
for several different input states. We set the qubits at positions 
$r_a$ and $r_b$ with 
coupling factors given by $g_{_\bk}^n=g_{_\bk} \e^{-i{\mbox{\small\boldmath$k$}} 
\cdot \br_n}$, and $n=a,b$. 
It is easy to see that the unitary evolution operator induces 
entanglement between
qubit states and field states: $U_I(t)$ acts as a conditional 
displacement operator for the field with a displacement amplitude
depending on both qubits of the QR, as discussed in more detail in 
\cite{palma}. As we have 
pointed out previously, it is this entanglement that is responsible 
for the decoherence processes 
described in the present paper. In particular, the case of 
two qubits has
$\rho_{_{i_aj_a,i_bj_b}}^Q(t)  = 
\bra i_{_a},i_{_{b}} | Tr_{_B}\{\rho^S (t)\}| j_{_a},j_{_{b}}\ket$. 
Next we analyze the register dynamics for the limiting decoherence situations 
described above. First, let us study the case of {\it independent decoherence}:
\begin{itemize}
\item[\ (i)] $i_a=j_a$, $i_b\neq j_b$: \\
$\rho_{_{i_ai_a,i_bj_b}}^Q(t)  =  
\exp\big[i \Theta_d(t)f_{i_ai_a,i_bj_b} -\Gamma_d(t;T)\big]\times
\rho_{_{i_ai_a,i_bj_b}}^Q(0)$, where $f_{i_ai_a,i_bj_b}=2i_a(i_b-j_b)
\cos\nbk\cdot\nbr_{ab}$. Hence $f_{00,01}=f_{11,10}=\cos\nbk\cdot\nbr_{ab}$, 
and
$f_{00,10}=f_{11,01}=-\cos\nbk\cdot\nbr_{ab}$: $\rho_{_{i_ai_a,i_bj_b}}^Q(t) $ 
shows collective decay. This result is contrary to the one reported in Ref.
\cite{palma}.
\item[\ (ii)] $i_a=j_a$, $i_b = j_b$: \\
$\rho_{_{i_ai_a,i_bi_b}}^Q(t)  =   
\rho_{_{i_ai_a,i_bi_b}}^Q(0)$; as expected, the populations remain unaffected.
\item[\ (iii)] $i_a\neq j_a$, $i_b\neq j_b$: \\
$\rho_{_{i_aj_a,i_bj_b}}^Q(t)  =  
\exp\big [-\Gamma_d(t;T)h_{i_aj_a,i_bj_b}\big] 
\rho_{_{i_aj_a,i_bj_b}}^Q(0)$, where $h_{i_aj_a,i_bj_b}=2[1+(i_a-j_a)(i_b-j_b)
\cos\nbk\cdot\nbr_{ab}]$. Clearly,
$h_{10,10}=h_{01,01}=2[1+\cos\nbk\cdot\nbr_{ab}]$, and $h_{10,01}=h_{01,10}=
2[1-\cos\nbk\cdot\nbr_{ab}]$. Hence $\rho_{_{i_aj_a,i_bj_b}}^Q(t)$ also shows
collective decay.
\end{itemize}
\noindent
In the above cases, analytic expressions for the corresponding decoherence functions 
can be found. As in Section III (a), we shall consider 
two different surrounding environments. In the low-TE regime, the `Ohmic 
environment' ($d=1$) induces the following coherence decay (the high-TE 
requires numerical integration):
\beqa 
\nonumber
\rho_{_{d=1}}^{\pm}(t) & \approx & \exp\bigg[-\Gamma_1(t,T)\pm{ic_1}\bigg(\frac{1}{2}\arctan(\w_ct_-)- \\ &&
       \frac{1}{2}\arctan(\w_ct_{_+}) + 
       \frac{\w_c t}{1+\w_c^2t_s^2} \bigg)\bigg] \rho_{_{d=1}}^{\pm}(0)\ 
\eeqa
for  $i_a=j_a$, and $i_b\neq j_b$.  For brevity, we have dropped the subindices
of the reduced density matrix, and set $t_{_+}=t_s+t$, and $t_-=t_s-t$. For
$i_a\neq j_a$, and $i_b\neq j_b$ we obtain the result 

\beqa 
\nonumber
           \rho_{_{d=1}}^{\pm}(t) & \approx & \exp\bigg[-2\Gamma_1(t,T)\mp\\ &&
\nonumber
            2c_1
           \bigg(\frac{1}{4}\bigg\{2\ln\bigg(\frac{1+4\w_{_T}^2 t_s^2}{1 + \w_c^2 t_s^2}\bigg)+\\ &&
\nonumber
            \ln\bigg(\frac{1 +\w_c^2t_-^2}{1 +4\w_{_T}^2t_-^2}\bigg) + 
             \ln\bigg(\frac{1 +\w_c^2t_{+}^2}{1 +4\w_{_T}^2t_{+}^2}\bigg)\bigg\} - \\ &&
\nonumber
            \w_{_T}\Big\{2t_s\arctan(2\w_{_T}t_s)-t_-\arctan(2\w_{_T}t_-) - \\ &&
             t_{_+}\arctan(2\w_{_T}t_{_+})\Big\}\bigg)\bigg] \rho_{_{d=1}}^{\pm}(0)\ .
\eeqa

In Figs. 1 and 2 we have plotted the decay of two qubit coherence due
to the coupling to an environment of the Ohmic type (Eqs. (23)). Here, 
$\Gamma_1^{\pm}(t,T)$
are defined from Eqs. (23) as $\rho_{_{d=1}}^{\pm}(t) =  
\exp[-\Gamma_1^{\pm}(t,T)]\rho_{_{d=1}}^{+}(0)$. From these figures we can see the variations 
of the onset of decay when increasing the temperature. Figure (1) shows how the  departure 
of coherence from unity changes with $t_s$ (plot 1 (i)) whilst for high temperature 
(plot 1 (iii)) there is no variation 
with $t_s$ at all. 
In the limit of large $t_s$ (see Table 1), we recover
the onset of decay 
of Fig. 6 (Appendix A). Note the difference between the time scales of plots 1 (i) and (iii), and how an 
estimation of typical decoherence times strongly relies on the temperature of the
environment. Figure (2) shows how the coherence decay shown in Fig. 1 disappears for 
small $t_s$ values, i.e. for a given temperature, it is possible to find a $t_s$
from which there is no decoherence of the two qubit system. This interesting
behaviour occurs 
only for the density matrix 
elements $
\bra 10 | Tr_{_B}\{\rho^S (t)\}|01\ket = 
\bra 01 | Tr_{_B}\{\rho^S (t)\}|10\ket$.

In Table 1 we give some typical two qubit decoherence times $\tau_{dec}$ for an Ohmic
environment in terms of the temperature, coupling constants $c_1$, and $t_s$. In all of
the tables in this paper, we have  taken the beginning of the decoherence process to occur when the
reduced density matrix of the whole system exhibits a 2\% deviation from the initial
condition, i.e. when $\exp[-\Gamma_i(t\equiv\tau_{dec},T)]=0.98$. The tables have been
generated from the corresponding decoherence functions reported in this paper. Here, 
$t_f$ is defined from $\exp[-\Gamma_i(t=t_f,T)]=0.01$, i.e. the
difference between $t_f$  and $\tau_{dec}$ gives an estimate for the duration of the
decoherence process, say $t_{decay}$; and $\tau^\pm_{dec}$, and $t^\pm_{f}$
are evaluated from the respective decoherence functions $\Gamma_i^{\pm}(t,T)$, with $i=1,3$.  
In order to gain insight into some characteristic time-scales, 
consider for example the case of the solid state, where in many situations the noise
field  can be identified 
with the phonon field. Here the cut-off $\w_c$ can be immediately associated with
the Debye frequency
$\w_{_D}$.  
A typical Debye temperature $\Theta_D=100$ K has $\w_{_D}\equiv
\w_c\approx10^{13}$ s$^{-1}$, so
$\theta\equiv\w_{_T}/\w_c\approx10^{-2} \ T$. Hence
we
can see from Table 1 (Figs. 1 and 2) that for $c_1=0.25$, $\w_ct_s=0.5$, and $T=0.1$ K,
the decoherence process starts at
$\tau^+_{dec}\approx  23.5$ fs and $\tau^-_{dec}\approx  43.7$ fs, and it lasts
for $t^+_{decay}\approx 10.3$ ps (for $t^-_{decay}$, $\exp[-\Gamma^-_1(t\mapsto\infty,0.1$ K)]
saturates above 0.01).  If the strength of the coupling 
goes down to $c_1=0.01$ then  
$\tau^+_{dec}\approx  0.2$ ps, and 
$t^+_{decay}\approx 3.5$ ns. In this latter case, $\tau^-_{dec}$, and $t^-_{decay}$ are not reported since 
the coherence saturates to a value above 0.98.  From the data reported in Table 1
is clear that the weaker the coupling between the qubits and the environment, the longer the decoherence
times and the slower the decoherence process. In addition the higher the temperature, 
the faster the qubits decohere, as  expected intuitively.

The effect of the transit time becomes more evident from Table 1: for large $t_s$ values
there is no difference between $\tau^+_{dec}$ and $\tau^-_{dec}$ (also $t^+_{decay}\approx t^-_{decay}$).
This is because for large $t_s$, the contribution due to terms involving $t_s$  in Eq. (23)
is negligible, hence the reduced density matrix $\rho_{_{d=1}}^{\pm}(t)\equiv 
\rho_{_{d=1}}(t)\approx 
\exp[-2\Gamma_1(t,T)]$  and hence has a similar behaviour to the single qubit
case (Eq. (19)).  Hence for large $t_s$ (e.g. $t_s=10^4$), Figs. 1 and 2 resemble the onset of decay
of the single qubit case (see later Fig. 6) where there is 
no dependence on $t_s$. We note that $\exp[-\Gamma^+_1(t,T]$ 
($\exp[-\Gamma^-_1(t,T]$)
is the corresponding decay of the coherences $\rho_{_{10,10}}=\rho_{_{01,01}}$ 
($\rho_{_{10,01}}=\rho_{_{01,10}}$), hence $\rho^Q_{_{10,10}}(t)=\rho^Q_{_{10,01}}(t)$
for large $t_s$.

By contrast, the super-Ohmic $d=3$ field leads to 
the following coherence decay

\beqa 
\nonumber
           \rho_{_{d=3}}^{\pm}(t) & = & \exp\bigg[-\Gamma_3(t,T)\pm \\ &&
\nonumber
           ic_3\bigg(
           \frac{ \sin(2\arctan(\w_c t_-))}{2(1 +\w_c^2t_-^2)} -
           \frac{\sin(2\arctan(\w_c t_{_+}))}{2(1 +\w_c^2t_+^2)}+\\ &&
           \frac{2\w_ct \cos(3\arctan(\w_c t_s))}{(1 +\w_c^2t_s^2)^{3/2}}\bigg)\bigg]
           \rho_{_{d=3}}^{\pm}(0)\  
\eeqa
\noindent
for  $i_a=j_a$, and $i_b\neq j_b$, and

\vspace{0.1cm}
{\small
\begin{center}\begin{tabular}{c|c|c|c|c|c|c}
\hline
$c_1$ & $k_BT/\hbar\w_c$ & $\w_c t_{s}$ & $\w_c \tau^-_{dec}$ & $\w_c t^-_{f}$
& $\w_c \tau^+_{dec}$  &  $\w_c t^+_{f}$
\cr\hline\hline
0.25 & $10^{-3}$  &  0.5     & 0.436919  & saturates  & 0.235446   & 103.507    \cr 
0.25 & $10^{0}$   & 0.5      & 0.183755  & saturates  & 0.104119   & 2.05958    \cr
0.25 & $10^{-3}$  & $10^{4}$ & 0.290113  & 1279.63    & 0.290113   & 1279.64    \cr 
0.25 & $10^{0}$   & $10^{4}$ & 0.127778  & 3.45901    & 0.127778   & 3.45901    \cr
0.1  & $10^{-3}$  & 0.5      & 0.913573  & saturates  & 0.37654    & 2025.75    \cr
0.1  &$10^{0}$    & 0.5      & 0.303135  & saturates  & 0.16504    & 4.28334    \cr
0.1  & $10^{-3}$  & $10^{4}$ & 0.47316   & 5669.66    & 0.473159   & 5670.15    \cr
0.1  &$10^{0}$    & $10^{4}$ & 0.203549  & 7.86596    & 0.203549   & 7.86596    \cr
0.01 & $10^{-3}$  & 0.5      & saturates &  saturates & 1.45274    & 35004.7    \cr
0.01 & $10^{0}$   & 0.5      & saturates & saturates  & 0.538502   & 37.2732    \cr
0.01 & $10^{-3}$  & $10^{4}$ & 2.55738   &  saturates & 2.55738    & 40816.8    \cr
0.01 & $10^{0}$   & $10^{4}$ & 0.709492  & 73.8325    & 0.709492   & 73.8325
\cr\hline
\end{tabular}\end{center}}
{\noindent\footnotesize Table 1. Characteristic times for two-qubit independent decoherence 
$\tau^\pm_{dec}$ for $d=1$ dimensional density of states of the field. Different 
temperature, transit time, and coupling
strength values are considered. $i_a\neq j_a$, $i_b\neq j_b$ (see text).}

\beqa 
\nonumber
           \rho_{_{d=3}}^{\pm}(t) & = & \exp\bigg[-2\Gamma_3(t,T)\mp 2c_3\bigg(-\frac{1-\w_c^2t_s^2}
      {[1+\w_c^2t_s^2]^2}+\\ &&
\nonumber
           \frac{1-\w_c^2t_{+}^2}{2[1+\w_c^2t_{+}^2]^2}+
           \frac{1-\w_c^2t_-^2}{2[1+\w_c^2t_-^2]^2}+\\ &&
\nonumber
          \frac{\theta^2}{2}\bigg\{
    2\zeta(2,\theta-i\omega_{_T} t_s)+ 
    2\zeta(2,\theta+i\omega_{_T} t_s)- \\ &&
\nonumber
    \zeta(2,\theta+i\omega_{_T} t_{_+})- 
    \zeta(2,\theta-i\omega_{_T} t_{_+})- \\ &&
\nonumber
    \zeta(2,\theta+i\omega_{_T} t_-)- 
    \zeta(2,\theta-i\omega_{_T} t_-)\bigg\}\bigg)\bigg]
           \rho_{_{d=3}}^{\pm}(0)\  \\ &&
\eeqa 
for $i_a\neq j_a$, and $i_b\neq j_b$.
The results of Eqs. (24) and (25) are exact:
no approximations have been made in obtaining them. Therefore, they are valid for 
any temperature of the environment.

In Figs. 3 and 4 we have plotted the decay of two qubit coherence due
to the coupling to a reservoir with three-dimensional density of states (Eqs. (25)).  
Some estimates for $\tau_c$ and $t_{decay}$ have been
given in Table 2. The decoherence functions $\Gamma_3^{\pm}(t,T)$ are defined as $\rho_{_{d=3}}^{\pm}(t) =  
\exp[-\Gamma_3^{\pm}(t,T)]\rho_{_{d=1}}^{\pm}(0)$.

\noindent 
As we can see, there is a {\it non-monotonic}
behaviour for the decay of coherence for low temperature values, as can be seen from Figs. 
3 (i), and (ii). The decay given by the functions  $\Gamma_3^{+}(t,T)$ (Fig. 3) 
and $\Gamma_3^{-}(t,T)$ 
(Fig. 4) saturates to a particular value, 
which is fixed by the strength of the coupling and the temperature of the 
reservoir: the lower the temperature, the slower the decay, and the higher the residual coherence.
Some estimates for these saturation values ${\mbox{e}}^{-\Gamma^\pm_3(t^\pm_{f},T)}$ 
are shown in Table 2.  
From Fig. 4 it is possible to find small $t_s$ values for which the onset of decay 
does not change in time and coherence remains unaffected. 
This result is very different from the case of Fig. 3, where coherence either vanishes 
(at high temperatures) or saturates to a residual coherence value (at low temperatures). 
Also note that whilst nothing happens to the onset of decay of Fig. 4, 
the coherence decay is amplified 
in the case of Fig. 3, for small $t_s$ values. 
From Figs. 3 and 4 (Figs. (i), and (ii)) we see that there is saturation of the decay in 
the presence of a 
{\it non-trivial} coherence process. However, for high temperatures (Figs. 3 (iii) and 4 (iii))
there is a monotonic behaviour where no 
saturation occurs at all and the residual coherence vanishes. 

\noindent
We note that the non-monotonic behaviour reported here for the low
temperature regime is not just 
a characteristic of high
dimensionality fields: it also occurs  for the $d=1$ field, 
as can be seen from Figs. 1 (i) and 2 (i). 
We believe that this purely 
quantum-mechanical phenomenon is associated 
with an interplay between the 
quantum vacuum and thermal fluctuations of the system and the
quantum character of the field. The system undergoes a dynamics where the
environment manages to `hit back' at the qubits in such a way that 
the coherences then exhibit an effective revival before vanishing
(at high temperatures) or saturating to a residual value (at low temperatures).  
An essential feature of the model studied here
is the fact that the QR and bath are assumed to be initially uncorrelated. In future 
work we will analyze the behaviour of these
`recoherences' with more general initial conditions, where the initial state of the
combined system is allowed to contain some correlations between the bath and the QR (see
also Ref.
\cite{amir}).  Such studies will help to clarify the origins of this `recoherence
effect' and also the effectiveness of decoherence as a function of the initial
conditions. We note that there is a previous work by Hu {\it et al.}  
\cite{juan-pablo} 
where the study of quantum brownian motion in a general environment gives rise to a similar 
behaviour (although in a different context) to the one reported here for the `recoherences'. 
A more detailed analysis of the physical implications of this interesting 
behaviour of the coherence decay is intended to be  addressed elsewhere
\cite {reina}.

It is interesting that the dynamics of coherences 
depends so strongly on the strength of the coupling. It can be seen from 
Table 2 and Fig. 5 that
i) the coherences saturate (sat.) to a very high `residual value' and  show less than a $2\%$ decay 
for weak coupling  (e.g. $c_3=0.01$) and low temperatures ($T=0.1$ K). This result is to be compared with
the $d=1$ case where the coherences always vanish for long $t_s$; 
ii) even at high temperatures ($T=10^4$ K) where the coherences vanish, and weak coupling, we find the 
appearence of the `recoherence effect' discussed above. 
This makes more evident the role of the QR-bath 
coupling: recoherences should  appear only under certain conditions imposed by the strength 
$c_3$ (hence by the spectral density) and the temperature. These conditions can be 
directly obtained from the decoherence functions reported in this paper. Typical $\tau^\pm_{dec}$ 
times for this $d=3$ dimensional environment are given in Table 2. From these we can conclude that the
two QR decoherence time scales are shorter, and the decoherence process occurs
faster, than in the single qubit case.

\vspace{-0.2cm}
{\footnotesize
\begin{center}\begin{tabular}{c|c|c|c|c|c|c|c|c}
\hline
$c_3$ & $\w_{_T}/\w_c$ &  $\w_c t_{s}$ &  $\w_c \tau^+_{dec}$
&  $\w_c
t^+_{f}$  & ${\mbox{e}}^{-\Gamma_3^+(t^+_{f})}$  &
$\w_c \tau^-_{dec}$ & $\w_c t^-_{f}$ 
 &
${\mbox{e}}^{-\Gamma_3^-(t^-_{f})}$ 
\cr\hline\hline
0.25  & $10^{-3}$  & 0.5      & 0.1292   & sat.  & 0.477  & 0.10818 & sat.  &0.771 \cr 
0.25  & $10^{2}$   & 0.5      & 0.01338  & 0.20  & 0.01   & 0.01522 & 0.24  & 0.01\cr
0.25  & $10^{-3}$  & $10^{2}$ & 0.11738   & sat.  & 0.6065  & 0.11738  & sat.  &0.6065 \cr 
0.25  & $10^{2}$   & $10^{2}$ & 0.01421  & 0.22  & 0.01   & 0.01421  & 0.22  & 0.01\cr
0.01  & $10^{-3}$  & 0.5      & 0.79957  & sat.  & 0.971  & sat. & sat.     &0.989\cr
0.01  & $10^{2}$   & 0.5      & 0.066994 & 1.51  & 0.01   & 0.07645 & sat.  &0.449\cr
0.01  & $10^{-3}$  & $10^{2}$ & 9.7767   & sat. & 0.9802  & 9.7767 & sat. &0.9802\cr
0.01  & $10^{2}$   & $10^{2}$ & 0.07124 & sat.  & 0.01831 & 0.07124 & sat.  &0.01832
\cr\hline
\end{tabular}\end{center}}
\vspace{-0.1cm}
{\noindent\footnotesize Table 2. Characteristic times for two qubit independent decoherence 
$\tau^\pm_{dec}$ for $d=3$ dimensional density of states of field. $i_a\neq j_a$, $i_b\neq j_b$.}


We now analyze how the above results are affected when we consider the situation 
of {\it collective decoherence}, as given by Eq. (16). The reduced density
matrix for the two qubit system now reads:
$ \rho_{_{i_aj_a,i_bj_b}}^Q(t) = 
\exp\big\{i \Theta_{d}(t)\big[\big(\sum_{m=a}^bi_m\big)^2-\big(\sum_{m=a}^bj_m\big)^2\big]
\big\}\exp\big\{-\Gamma_d(t;T)\big[\sum_{m=a}^b(i_m-j_m)\big]^2\big\}\times 
\rho_{_{i_aj_a,i_bj_b}}^Q(0)$. 
In particular, we find:
\vspace{0.3cm} \\
\noindent
(i) $i_a=j_a$, $i_b\neq j_b$:  
\\
$\rho_{_{i_ai_a,i_bj_b}}^Q(t)  =  
\exp\big[i \Theta_d(t)f^{'}_{i_ai_a,i_bj_b}-\Gamma_d(t;T)\big]
\rho_{_{i_ai_a,i_bj_b}}^Q(0),
$ where $f^{'}_{i_ai_a,i_bj_b}=2i_a(i_b-j_b)$. Hence $f^{'}_{00,01}=f^{'}_{11,10}=1$, 
and
$f^{'}_{00,10}=f^{'}_{11,01}=-1$. The corresponding decoherence rates are
\beqa 
\nonumber
\rho_{_{d=1}}^{\pm}(t) & \approx & \exp\big[-\Gamma_1(t,T)\pm{ic_1}\big(\w_ct+
\arctan(\w_ct)\big)\big] \rho_{_{d=1}}^{\pm}(0)\ 
\eeqa
\vspace{-0.7cm}
\beqa
\nonumber
           \rho_{_{d=3}}^{\pm}(t) & = & \exp\bigg[-\Gamma_3(t,T)\pm \\ &&
           ic_3\bigg(2\w_ct-
           \frac{ \sin(2\arctan(\w_c t))}{1 +\w_c^2t^2}\bigg)\bigg]
           \rho_{_{d=3}}^{\pm}(0)\  
\eeqa
for the Ohmic environment and the $d=3-$dimensional density of states, respectively.
\vspace{0.3cm} \\
\noindent
(ii) $i_a=j_a$, $i_b = j_b$:  
\\
\noindent
$\rho_{_{i_ai_a,i_bi_b}}^Q(t)  =   
\rho_{_{i_ai_a,i_bi_b}}^Q(0)$: as expected, the populations remain unaffected.
\vspace{0.3cm} 
\\
(iii) $i_a\neq j_a$, $i_b\neq j_b$: 
\\
\noindent
$\rho_{_{i_aj_a,i_bj_b}}^Q(t)  =  
\exp\big [-\Gamma_d(t;T)h^{'}_{i_aj_a,i_bj_b}\big] 
\rho_{_{i_aj_a,i_bj_b}}^Q(0),
$ where $h^{'}_{i_aj_a,i_bj_b}=(i_a+i_b-j_a-j_b)^2$. Then,
$h^{'}_{10,10}=h^{'}_{01,01}=4$, and $h^{'}_{10,01}=h^{'}_{01,10}=0$. Obviously, 
the corresponding decoherence rates are:

\beqa 
\nonumber
            \rho_{_{d=1}}^{-}(t) & = & \rho_{_{d=1}}^{-}(0)\ ,  \\
            \rho_{_{d=1}}^{+}(t) & \approx & \exp\big[-4\Gamma_1(t,T)\big] \rho_{_{d=1}}^{+}(0)\ 
\eeqa
for the $d=1$ dimensional field, and
\beqa 
\nonumber
          \rho_{_{d=3}}^{-}(t) & = & \rho_{_{d=3}}^{-}(0)\ , \\
         \rho_{_{d=3}}^{+}(t) & \approx & \exp\big[-4\Gamma_3(t,T)\big] \rho_{_{d=3}}^{+}(0)\ 
\eeqa 
for the $d=3$ dimensional field. Hence, regarding the input states, the case of 
{\it collective decoherence} shows two very well defined situations: \vspace{0.2cm}\\
(a) A set of input states that shows no
decoherence at all, despite the fact that the qubits are interacting with the 
environment.
This is because  under the specific situation of collective coupling there is a set
of  initial qubit states that does not 
entangle with the bosonic field and hence the states preserve their coherence. 
These states are the so-called ``coherence-preserving'' states and, for the case of
a $2-$QR, the corresponding reduced density matrix elements are
$\bra 10 | Tr_{_B}\{\rho^S (t)\}| 01\ket$, and $\bra 01 | Tr_{_B}\{\rho^S (t)\}
|10\ket$. As studied in more detail in Ref. \cite{dec-free} (where relaxation 
effects are also included), this result can be used as 
an encoding strategy, 
where an arbitrary $L-$QR 
can be decoupled from its environment provided that every single qubit of the register 
can be encoded using 2 qubits: e.g. using the simple encoding (though not the 
most efficient one)
$|0\ket\mapsto |01\ket$, and
$|1\ket\mapsto |10\ket$. It turns out that this procedure is ensured even if 
relaxation effects are included in the Hamiltonian (1) \cite{dec-free}.
Hence in the specific situation of collective decoherence one can find, for arbitrary 
$L$,  a decoherence-free  subspace (DFS) $C_{_L}\in {\cal H}^{\otimes L}$
(the Hilbert space ${\cal H}={\cal H}_{QR}\otimes {\cal H}_B$) that does not get 
entangled with the 
environment, and hence
the QR should evolve without decoherence. 
Besides quantum error correction codes, this is currently one of the most outstanding 
results in the battle against decoherence 
\cite{dec-free}, particularly because of its relevance to maintaining a coherent 
qubit memory in quantum information processing. \vspace{0.2cm}\\
(b) The other two input states have the decay of coherences $\bra
11 | Tr_{_B}\{\rho^S (t)\}| 00\ket$ and 
$\bra 00 | Tr_{_B}\{\rho^S (t)\}|11\ket$: these give a situation 
of `superdecoherence' \cite{palma}, where the qubits are collectively entangled 
and hence these matrix elements give the fastest decay for the coherences. 
This means that whilst there is a subspace which is robust against decoherence as
discussed above,
the remaining part of the Hilbert space gets strongly entangled with the environment.
This superdecoherence situation is illustrated in Fig. 6 for the case of reservoirs
with one and three dimensional-density of states. 

\noindent
The above process of calculating explicit results for the decay of any coherence
associated with
the coupling of a $L-$QR to a bosonic reservoir, for both types of 
coupling (independent and collective), 
can be carried out for any $L>2$  using the general
formulas Eqs. (12) and (16). In so doing, we can obtain an estimate of
typical decoherence times for a QR with an arbitrary number of qubits.
We should point out that if no schemes for controlling 
the errors induced by  the decoherence phenomenon  are
used 
\cite{shor2,knill,dec-free,using-dec,julia,lorenza1,suppression}, 
$\tau_{dec}$ establishes  an upper bound to the duration 
of any reliable quantum computing process.


\section{Discussions and concluding remarks}

We have revisited 
a model of decoherence based upon the one 
previously studied
by Leggett et al. \cite{leggett} in connection with the tunneling problem in 
the presence of dissipation, and  used later by Unruh and Palma {\it et al.}
\cite{unruh,palma} for  describing the decoherence process of a quantum register.
We have presented here a more complete description of this latter problem, which
provides the following new results. The decoherence rates of the density matrix
elements are correctly derived, leading to new quantitative results for both
independent  and collective decoherence situations. 
As
discussed here, if no error correcting/preventing
strategies are used, this has  implications for an estimation of decoherence time-scales
for which the  quantum memory of a QR
can be maintained in any reliable computation (we note, however, there has been a recent 
proposal by Beige {\it et al.} regarding the use of dissipation to remain and manipulate 
states within a DFS \cite{using-dec}). 
Our results agree with
those reported in \cite{duan}	for the case of collective decoherence but they
are different to the ones reported there for the case of independent decoherence:
in Ref.  \cite{duan} it is argued that independent decoherence, as opposed to 
collective decoherence, is insensitive to the qubit input states. Here we have shown 
instead that both cases are very sensitive to the input states and that both of them show 
collective decay. 

In the specific situation of independent decoherence we have found a non-trivial 
behaviour  in the decay of the off-diagonal elements of the reduced density matrix. Here 
the coherences experience an effective revival before they either vanish
or saturate to a residual value. This 
behaviour depends on the temperature of the environment and depends strongly on the strength 
of the coupling: the coherence dynamics are different in the weak and strong coupling regime.
Also, there are important qualitative differences between the Ohmic $d=1$ and the super-Ohmic 
environment $d=3$, which is ultimately linked to the spectral density of the bath. In the former case
the coherence is always lost, whilst in the latter the coherence generally saturates to a 
residual value which is fixed by the temperature and strength of the coupling 
and $only$ vanishes in the high-TE.
By contrast, in the case of collective decoherence we have identified QR input states 
that allow the system to evolve in a decoherence-free
fashion, the so-called coherence-preserving states. 
We note that DFS's do not exist in 
the specific situation of independent qubit decoherence. We also note that our 
attention has been drawn to
a dynamical-algebraic description that unifies the currently known quantum errors 
stabilization procedures \cite{paolo2} (see also Refs. 
\cite{lorenza1}(a), \cite{lorenza2}). 
Within the framework of a system formed by a 
collection of $L$ uncorrelated clusters 
of subsystems where 
each cluster fulfills the requirements of collective decoherence, 
Zanardi has shown that noiseless subsystems can be built \cite{paolo2}.

From the point of view of complexity analysis (and without including any strategy 
for the stabilizing of quantum information), we should ask how the results
reported in  the present paper affect those of Ref. \cite{palma}. We must 
identify the coherences that are destroyed more rapidly: from Eqs. (12) and
(16) 
it is easy to see that the coherences with the fastest decay are given by the
matrix elements 
$\rho_{_{\{0_n,1_n\}}}$ and $\rho_{_{\{1_n,0_n\}}}$. These off-diagonal
elements decay as  
$\exp[-\Gamma_d(t;T)f(L)]$, with
\beqa
f(L)=L+2\sum_{m=1,n=2 \atop m\neq
n}^L(i_m-j_m)(i_n-j_n)\cos(\w t_s) \ ,
\eeqa
for the limit of 
independent qubit decoherence,
and as $\exp[-L^2\Gamma_d(t;T)]$ for the collective decoherence case. Hence,
it is clear
that for both cases, the longer the QR coherence length, the faster the
coherence decay.
Despite the fact that the results of Palma et
al. \cite{palma} are not the same as ours, it turns out that both sets of
results lead to the same 
unwelcome exponential increase of the error rate. We note that 
the result of Eq. (29) is in general different to the one reported in 
\cite{palma}. We also note that the coherence decay  for the case
of collective decoherence coincides with that
of \cite{palma} only for the fastest off-diagonal 
element decay: if we consider different 
density matrix elements, the results of Ref. \cite{palma} no longer coincide with ours. 
If the
information  reported in our work is used for an estimation of the 
actual decoherence time associated with any given off-diagonal density
matrix element (coherence), 
the results are in general quite different
from the ones reported in \cite{palma}.
 
We have shown how a bosonic environment destroys
the coherences 
of an arbitrary quantum register. In doing so, we have identified 
DFS states that are 
invariant under the coupling 
to such an environment. This result could be of crucial importance 
for improving
the efficiency of 
quantum algorithms, for example. We believe that the engineering of DFS's 
will become intrinsic to
the designs of future quantum  computation architectures. There was a recent experimental
demonstration of decoherence-free quantum memories
\cite{kielpinski,kwiat}. This has been 
achieved for $one$ qubit, by encoding it into the DFS of a pair of trapped
$^9Be^+$ ions: 
in this way, Kielpinski et al. have demonstrated  the immunity of a DFS of
two atoms 
to collective dephasing \cite{kielpinski}. Prior to this experiment with
trapped ions, Kwiat et al. demonstrated the robustness of a DFS for two
photons to collective noise \cite{kwiat}. 
Robust quantum memories seem therefore to be well on their way, both theoretically and 
experimentally, to overcoming the main obstacle 
to quantum information processing - namely, decoherence.

\vspace{0.5cm}
\noindent {\bf Acknowledgements.} 
We are grateful to L. Viola for a critical reading of the
manuscript.  J.H.R. acknowledges H. Steers for continuous
encouragement and G. Hechenblaikner for stimulating discussions. J.H.R. is grateful
for financial suppport  from the Colombian  government agency for science and 
technology (COLCIENCIAS). L.Q. was partially supported by COLCIENCIAS under contract
1204-05-10326. N.F.J. thanks EPSRC for support.

\appendix


\section{Single qubit decoherence}

The
decoherence rates for a single qubit coupled to
a reservoir with $d=1$, and $d=3$ density of states are
(Eq. (19))
\beq 
\Gamma_1(t,T)  = \frac{\alpha_1}{\hbar^2} \int d\w \e^{-\w/\w_c} 
\frac{1-\cos({\w}t)}{{\w}}\coth\Big(\frac{\w}{2\w_{_T}}\Big)
\eeq
\vspace{-0.3cm}
\beq 
\Gamma_3(t,T)  = \frac{\alpha_3}{\hbar^2}\int d\w \ \w \e^{-\w/\w_c}[1-\cos({\w}t)]
\coth\Big(\frac{\w}{2\w_{_T}}\Big).
\eeq
In Section III A we gave the analytic solutions to these integrals. However, we did not 
perform a full analysis of those results. We start here by recalling that the 
solution found for the integral (A1)
was an approximate one, valid only for the low-TE ($\w_{_T}<<\w_c$): a general solution to
this integral requires numerical integration. The second integral was solved analytically 
for any temperature value $T$ (without making any approximation).
In these calculations
note that the constant coupling $\alpha_d$ changes its units with the 
dimensionality of the field: $[\alpha_1]=$[(eV)$^2$s$^2$],
$[\alpha_3]=$[(eV)$^2$s$^4$], etc. 

We first analyze the case $d=1$. In the low-TE,   
Eq. (20) leads to the identification of three main regimes for the 
decay of the coherences:  (a) a ``quiet" regime, for which $t<\tau_c$, and
$\Gamma_1(t,T) 
\approx c_1\w_c^2t^2/2$; (b)  a ``quantum'' regime, where $\tau_c <t <\tau_{_T}$, and
$\Gamma_1(t,T) 
\approx c_1\ln(\w_ct)$; and (c) a ``thermal'' regime, for which $t >> \tau_{_T}$, 
and  $\Gamma_1(t,T)  \approx
2c_1\omega_{_T}t$. These regimes have also been discussed in Refs. \cite{palma,unruh} 
and can be easily identified in Fig. 7 for several different 
temperatures. 
In Fig. 7 (i) we have plotted Eq. (19) as a
function of $\w_ct$ for several different temperatures and for $d=1$. 
Since the decoherence effects arising from 
thermal noise
can be separated from the ones due to quantum vacuum fluctuations, we have 
also plotted these partial contributions in order to see their effects over 
the time-scales involved in the decoherence of the single qubit (Eq. (A1)). 
It can be seen that for a given value of the temperature  parameter $\theta$,
a characteristic time for which 
we start observing deviation of coherence 
from unity is determined by the shortest of the two time-scales $\tau_c$ and $\tau_{_T}$,
and that this value is increased  
when decreasing the temperature $T$, as expected. 

\noindent
From Fig. 7 (i.c) it can clearly be seen that at low temperatures the quantum 
vacuum fluctuations play the major role in eroding the qubit coherence
whilst the contribution due to thermal fluctuations plays a minor role. From
this plot we can 
see the three main regimes indicated above: a quiet ($t<\tau_c$), a quantum 
($\tau_c < t < \tau_{_T}$) and a thermal ($t > \tau_{_T}$) regime. 
In this limit of low-TE, $\tau_c$ is the  characteristic time that 
signals the departure 
of coherence from unity. Here the qubit dynamics shows a competition between
contributions arising from vacuum and thermal fluctuations: even at thermal 
time scales, the contribution to the decoherence due to vacuum fluctuations remains
important.

In the case of the high-TE, the decay
due to thermal noise (see dashed line in Fig. 7 (i.a)) becomes more important 
than the vacuum fluctuation  contribution\cite{note1} and
the start of the decoherence process is ruled by $\tau_{_T}$. 
Similar conclusions can be 
obtained from Fig. 7 (ii), where the decay of coherence 
has been plotted as a function of time but  in units of the thermal frequency
$\w_{_T}$ for several different temperatures.

We
can see from Table 3 and Fig. 7 that $t_{decay}$ is comparable to $\tau_c$ for the
high-TE, and to
$\tau_{_T}$ for the low-TE. Indeed, if we assume $\w_c$ to be the Debye cut-off
($\w_c\sim 10^{13}$ s$^{-1}$), we obtain from Table 3 that for $c_1=0.25$, and $T=1$ mK,
the decoherence process starts at
$\tau_{dec}\approx  41.9$ fs and lasts
for
$t_{decay}\approx 27.4$ ns (for $d=1$). Here
$\w_{_T}=1.3\times 10^{11}\ T
\approx 1.3\times 10^{8}$ s$^{-1}$, hence $\tau_{_T}\sim 8 $ ns is of the same
order of magnitude as $t_{decay}$. For the super-Ohmic environment
$d=3$  ($c_3$ and
$T$ as before) we obtain
$\tau_{dec}\approx 16.8$ fs. In this case the coherences saturate to a residual value, 
as we discuss below (see Table 3 and Fig. 8 (ii)). From Table 3
we can also see the effects of the high-TE: the qubit decoheres several orders of
magnitude faster than in the low-TE case. For example, for $T=100$ K, 
$k_BT>\hbar\w_c$ ($\hbar\w_c=6.58$ meV), and we obtain  $\tau_{dec}\approx  18.1$ fs,
and
$t_{decay}\approx 0.6$ ps ($c_1=0.25$, $d=1$). A similar behaviour is also observed in
the
$d=3$ case. 

\vspace{-0.2cm}
{\small
\begin{center}\begin{tabular}{c|cccccc}
\hline
$d$ & $c_i$ & $\w_{_T}/\w_c$ & $\w_c \tau_{dec}$
& ${\mbox{e}}^{-\Gamma_i(\tau_{dec})}$ & $\w_c t_{f}$ & 
${\mbox{e}}^{-\Gamma_i(t_{f})}$ 
\cr\hline\hline
& 0.25 & $10^{-5}$ & 0.418831 & 0.98 &
273950.34 & 0.01  \cr 
 & 0.25 & 1.0 & 0.181611 & 0.98 & 6.39891 & 0.01 \cr
$d=1$ & 0.1 & $10^{-5}$ &
0.705612 & 0.98 & 1153307.91 & 0.01  \cr
 & 0.1 & $1.0$ & 0.291365 & 0.98 & 15.19703 &
0.01 \cr
& 0.01 & $10^{-5}$ &
7.47367 & 0.98 & 14346140.39  & 0.01 \cr
& 0.01 & 1.0 &
1.09604 & 0.98 & 147.12606  & 0.01
\cr\hline\hline
 & 0.25 & $10^{-5}$ & 0.167969 & 0.98 & saturates 
& 0.778801 
\cr
& 0.25 & 1.0 & 0.154762 & 0.98 & saturates &
0.564132  \cr  
 & 0.25 & $10^{2}$ & 0.020104 & 0.98 & 0.318417 &
0.01 \cr
$d=3$ & 0.1 & $10^{-5}$ & 0.275766 &
0.98 & saturates & 0.904837  \cr 
& 0.1 & $1.0$ & 0.251550 & 0.98 & saturates & 0.795339 
\cr  
 & 0.1 & $10^{2}$ & 0.031791 & 0.98 & 0.546769 &
0.01  \cr 
 & 0.01 & $10^{2}$ & 0.101012 & 0.98 & saturates &

0.135331
\cr\hline
\end{tabular}\end{center}}
{\noindent\footnotesize Table 3. Single qubit decoherence times for
different temperatures, and coupling  strength $c_i$ ($i=1,3$); for $d=1$ (Ohmic), and
$d=3$ (super-Ohmic) dimensional density of states of the field.}

\vspace{0.4cm}
\noindent
In Table 3 we can appreciate the effect of the coupling strength on
the decoherence time scales. Let $c_1=0.01$  ($d=1$), hence  for i) $T=1$ mK, 
$\tau_{dec}\approx  0.75$ ps and
$t_{decay}\approx 1.4 \ \mu$s; for ii) $T=100$ K, 
$\tau_{dec}\approx  0.11$ ps and
$t_{decay}\approx 14.6$ ps.
Hence the weaker the coupling between the qubit and the environment, 
the longer the decoherence
times and the slower the duration of the decoherence process.
This result also holds for the case $d=3$, as can be seen from Table 3. All of
the above analysis concerning  the different regimes for the decay of the coherences
presented here is explicitly  illustrated in the three-dimensional plots of Fig. 8. 

Next, we analyze the decoherence behaviour of the single qubit when 
coupling to the super-Ohmic $d=3$ reservoir. The corresponding decoherence 
function is given by Eq. (21).
As  can be seen from Fig. 8 (ii) and Table 3, this case shows an interesting  
behaviour for the coherence
decay: once the end of the `quiet' regime has been reached, 
the coherences decay to either zero, as in the case of Fig. 8 (i), or 
saturate to a particular value
determined by the temperature parameter $\w_{_T}/\w_c$. Here we can 
identify the particular 
temperature value for which no saturation occurs at all
and the expected decoherence takes place. In Table 3 we give some saturation
values for different temperatures and coupling strengths $c_3$. For 
$c_3=0.1$, and $0.25$ the temperature value for which any residual
coherence vanishes falls in the interval $10<\theta<100$ (high-TE). Apparently these
residual values shown in Fig. 8 (ii) and Table 3 vanish when additional
frequency modes associated with the three-dimensionality of the field are taken into
account \cite{palma}. Even if this is not the case, surely the effects of relaxation 
mechanisms  would mark the extent of these 
residual values.


\section{Time evolution and the reduced density matrix}

In this Appendix we give details of the main steps followed in the 
calculation of the reduced density matrix given by Eq. (15). 

\subsection{The time evolution operator $U_I(t)$}

Since Eq. (3) gives $H_{_I}(t)=\sum_{n,\bk}  J_z^n \big(g_{\bk}^n \e^{i \wk t} 
b_{\bk}^\dag + g_{\bk}^{*n} \e^{-i \wk t} b_{\bk} \big)$, we have that

\beqa
\nonumber
U_{_I}(t) & = & T \exp \Big[ -\frac{i}{\hbar}\int_o^t  
\sum_{\bk}  g_{\bk}\big(\e^{i \wk t'} J_z^\bk \  b_{\bk}^\dag  \ + 
\\ && 
\e^{-i \wk t'} J_z^{\dag \bk} \ b_{\bk} \big) dt'\Big],
\eeqa
where we have introduced the shorthand notation 
$J_z^\bk=\sum_{n} \e^{-i{\mbox{\small\boldmath$k$}} 
\cdot \br_n} J_z^n$. 
Hence, we can rewrite Eq. (B1) as \cite{mahan}:
\beqa
\nonumber
U_{_I}(t) & = &  \exp \Big[\sum_{\bk} g_{_\bk}\varphi_{\w_{_k}}(t) 
J_z^\bk \  b_{\bk}^\dag \Big]\times  \\ &&
\nonumber
T \exp\Big[ -\frac{i}{\hbar}\int_o^{t'} dt'  \sum_{\bk} \e^{-i \wk t'} \times \\ &&
\nonumber
\exp\Big({-\sum_{\bkp}g_{_\bkp}\varphi_{\w_{k'}}(t') J_z^\bkp \  
b_{\bkp}^\dag}\Big)
g_{_\bk} J_z^{\dag\bk} \ b_{\bk}\times \\ &&
\exp\Big({\sum_{\bkp}g_{_\bkp}\varphi_{\w_{k'}}(t') J_z^\bkp \  
b_{\bkp}^\dag}\Big)\Big] \ .
\eeqa
It is easy to show that the calculation of the product given by 
the last two lines Eq. (B2) gives the result
\beq
g_{_\bk} J_z^{\dag\bk} \Big[\ b_{\bk}+g_{_\bk}\varphi_{\wk}(t) 
J_z^\bk\Big] \ .
\eeq
Hence, the following expression for $U_{_I}(t)$ arises
\beqa
\nonumber
U_{_I}(t) & = & {\mbox{\large e}}^{^{\sum_{\bk} g_{_\bk}\varphi_{\wk}(t) J_z^\bk b_{\bk}^\dag}} \
{\mbox{\large e}}^{^{-\sum_{\bk}g_{_\bk}\varphi_{\wk}^\ast(t) J_z^{\dag\bk}  
b_{_\bk}}}\times \\ &&
\nonumber
\exp\Big\{-\frac{i}{\hbar} \sum_{\bk} |g_{_\bk}J_z^\bk|^2 \int_o^{t} dt'
\varphi_{\wk}(t') \e^{-i \wk t'}\Big\} = \\ &&
\nonumber
\exp\Big\{\sum_{\bk} g_{_\bk}\big[\varphi_{\w_{_k}}(t) J_z^\bk \  
b_{\bk}^\dag - {\varphi_{\wk}^\ast(t) J_z^{\dag\bk} \  
b_{\bk}\big]}\Big\}\times \\ &&
\nonumber
\exp\bigg\{\sum_{\bk} \big|g_{_\bk}J_z^\bk\big|^2\bigg
[\frac{it}{\hbar^2{\wk}}
-\frac{\varphi_{\wk}^\ast(t)}{\hbar{\wk}}+\frac{|\varphi_
{\wk}(t)|^2}{2}\bigg]\bigg\}
\eeqa
\vspace{-0.5cm}
\beqa
\eeqa
where we have used the result 
$
\e^{A+B}=\e^{A}\e^{B}\e^{-[A,B]/2},
$
which holds for any pair of operators
$A$,
$B$ that satisfy
$[A,[A,B]]=0=[B,[A,B]]$ (as in the case of Eq. (B4)).  
It is
straightforward to see that Eq. (B4) gives the final result
\beqa
\nonumber
U_{_I}(t) & = & 
\exp\Big[i\sum_{\bk}\big|g_{_\bk}\big|^2 \ 
\frac{\w_{_k}t-\sin({\w_{_k}}t) }{(\hbar{\w_{_k}})^2} 
J_z^{\dag\bk} J_z^\bk\Big]\times \\ &&
\exp \Big [\sum_{\bk} 
\left\{ A_{\bk}^\dag(t)  - A_{\bk}(t) \right\}
\Big] \ ,
\eeqa
where $A_{\bk}(t)= g_{_\bk}^{\ast}\varphi_{\wk}^\ast(t) 
J_z^{\dag\bk} \ b_{\bk} $. 
 
\subsection{The reduced density matrix of a $L-$qubit register}

We start by using 
the result for $U_I(t)$ in order to calculate the decay of 
the coherences, i.e. $Tr_{_B}\big[\rho^B (0)U^{\dag
\{j_{_n}\}}_{_I}(t)U^{\{i_{_n}\}}_{_I}(t)\big]$, with  
$U_I^{\{i_n\}}(t)$ as  defined in Eq. (10). In so doing, we first compute 
the operator algebra for
the product
$U^{\dag\{j_{_n}\}}_{_I}(t)U^{\{i_{_n}\}}_{_I}(t)$ by taking into 
account the expression (B5). The result gives 
\beqa 
\nonumber
U^{\dag\{j_{_n}\}}_{_I}(t)U^{\{i_{_n}\}}_{_I}(t)  & = & 
\exp\Big[i \sum_{\bk}\big|g_{_\bk}\big|^2 \ 
\frac{\w_{_k}t-\sin({\w_{_k}}t) }{(\hbar{\w_{_k}})^2}\times \\ &&
\nonumber
\eeqa
\vspace{-0.9cm}
\beqa
\nonumber
\sum_{m,n}(i_mi_n-j_mj_n)\cos\nbk\cdot\nbrmn\Big]
\exp \Big [i\sum_{\bk}\big|g_{_\bk}\varphi_{\wk}(t)\big|^2 \times
\eeqa
\vspace{-0.5cm}
\beqa
\sum_{m,n}i_mj_n\sin\nbk\cdot\nbrmn\Big]\exp\Big [\sum_\bk\Big
(\sigma_{\bk}b_{\bk}^\dag-\sigma_{\bk}^\ast b_{\bk}\Big) \Big] \ , 
\eeqa
where we have set $\sigma_{\bk}\equiv 
g_{_\bk}\varphi_{\wk}(t)\sum_m(i_m-j_m)\e^{-i {\mbox{\small\boldmath$k$}}\cdot \br_m}$. 
From the above equation 
note that 
the first two exponential terms commute, hence we only have to take the trace over 
the third term. By doing this 
(see e.g. Ref. \cite{gardiner}), we obtain the
result
 \beqa
\nonumber
\hspace{-1cm}
Tr_{_B}\Big[\rho^B (0)\exp\Big\{\sum_\bk\Big(\sigma_{\bk}b_{\bk}^\dag-
\sigma_{\bk}^\ast b_{\bk}\Big) \Big\} \Big] = 
\eeqa
\nonumber
\vspace{-0.5cm}
\beqa
\nonumber
\hspace{-.3cm}
\prod_\bk\exp\Big [-\big|g_{_\bk}\big|^2 \ 
\frac{1-\cos({\w_{_k}}t)}{(\hbar{\w_{_k}})^2}\coth\bigg
(\frac{\hbar\omega_{_k}}{2k_{_B}T}
\bigg)\times
\eeqa
\vspace{-0.5cm}
\beqa
\hspace{-1.0cm}
\sum_{m,n}(i_m-j_m)(i_n-j_n)\cos\nbk\cdot\nbrmn\Big] \ , 
\eeqa
from where Eq. (11) arises immediately.

\newpage
\bigskip 
{\small
\centerline{\bf Figure Captions} 
\bigskip 
Figure 1. Two qubit ``independent decoherence" due to the coupling to a reservoir of 
the Ohmic type ($d=1$) as a function of time $t$ and the 
transit time $t_s$, for the input states associated with $\Gamma_1^{+}(t,T)$
($i_a\neq j_a$, $i_b\neq j_b$) . Here
$c_1=0.25$, and (i) $\theta=10^{-3}$, (ii) $10^0$, and (iii) $10^2$.
$\Gamma_i^{\pm}(t,T)$, with $i=1,3$ are defined in the text.

Figure 2. Two qubit ``independent decoherence" caused by the coupling 
to an `Ohmic  
environment' as a function of times $t$ and $t_s$, 
for the input states associated 
with $\Gamma_1^{-}(t,T)$ 
($i_a\neq j_a$, $i_b\neq j_b$). 
$c_1=0.25$, and (i) $\theta=10^{-3}$, 
(ii) $10^0$, and (iii) $10^2$.

Figure 3. Two qubit ``independent decoherence" due to the 
super-Ohmic environment $d=3$ (Eq. (25)) as a function of times 
$t$ and $t_s$, for the input states associated with $\Gamma_3^{+}(t,T)$. 
$c_3=0.25$, and (i) $\theta=10^{-3}$, (ii) $10^0$, and (iii)
$10^2$. 

Figure 4. Two qubit ``independent decoherence" due to the super-Ohmic  
environment $d=3$ as a function of times $t$ and 
$t_s$ for the input states associated with $\Gamma_3^{-}(t,T)$. 
$c_3=0.25$, and (i) $\theta=10^{-3}$, (ii) $10^0$, and
(iii) $10^2$. 

Figure 5. Two qubit ``independent decoherence". 
$d=3$, coupling strength $c_3=0.01$, and (i), 
(iii) $\theta=10^{-3}$, and (ii), (iv) $\theta=10^2$.

Figure 6. Two qubit ``collective decoherence" for (i) $d=1$ 
`Ohmic environment'  
(Eq. (27)), and (ii) $d=3$ `super-Ohmic environment'  
(Eq. (28)), as a function of time and the temperature 
$\theta\equiv \w_{_T}/\w_c$.
$c_1=c_3=0.25$, and $i_a\neq j_a$, and $i_b\neq j_b$.  
$\Gamma_{_{d}}^{+}(t,T)$ ($d=1,3$) is defined using Eqs. (27) and (28). 

Figure 7. (i) Decoherence of a single qubit for an `Ohmic environment' 
as a function of $t$ (in units of $\w_c$). The contributions arising from 
the separate integration of thermal ($\exp[-\Gamma_{_T}(t)]$) and vacuum 
($\exp[-\Gamma_{_V}(t)]$)
fluctuations are shown as dotted curves. 
$c_1=0.25$, (a) $\theta\equiv \w_{_T}/\w_c=1$, (b) $10^{-2}$, 
(c) $10^{-5}$. If $\w_c$ is the Debye cutoff, $\theta\approx10^{-2}\ T$  
(see text): the decoherence shown corresponds to $T=$ 100 K,
$T=1$ K, and $T=1$ mK,  
respectively. (ii) Coherence decay for (a) $\theta=10^{-5}$, (b) $10^{-2}$, (c) $10^{2}$. 
$c_1=0.25$. Here time is given in units of the thermal frequency $\w_{_T}\equiv k_{_B}T/\hbar$. 

Figure 8. Decoherence of a single qubit for (i)  $d=1$, and (ii) $d=3$, as a function of time 
(in units of the cut-off $\w_c$) and3
the temperature parameter $\theta\equiv \w_{_T}/\w_c$. 
$c_1=c_3=0.25$.  If $\w_c$ is the Debye cut-off, the range of coherence 
decay goes from a few mK up to (a) $10^4$ K (plot (i)) and (b) $1.5\times10^3$ K (plot (ii)).
}


\begin{references}
\bibitem{nota0} If relaxation mechanisms are also considered, in general
$J^n_x,J^n_y\neq 0$.
The presence of 
distinct error generators along $x, y, \mbox{and}\ z$ make the noise processes 
non-Abelian and the Hamiltonian can no longer be solved  analytically.
\bibitem{nota1} This factorizable condition is the one that has been
extensively studied so far. However, more general situations have been analyzed in
connection with the influence functional path-integral method of Feynman
and Vernon. See, e.g.,  Ref. \cite{amir}.
\bibitem{nota2} Formally, noise processes can be characterized in terms of the
interaction algebra that they generate, see e.g. Refs. \cite{lorenza2,paolo2}.
\bibitem{note1} Over the time-scale considered  for curve (a) in
Fig.  7 (i), the vacuum fluctuation  plot overlaps with curve (b), obscuring 
it from view. The same happens with the  vacuum fluctuation contribution associated with
curve (b): this gets superposed with curve (c) and neither can be directly observed in
Fig. 7 (i).
\end{references}
\end{document}